\documentclass[12pt,a4paper]{article}

\usepackage{lineno}  

\usepackage{graphicx}  

\usepackage{xspace}
\usepackage{color}
\usepackage{colortbl}

\usepackage{amsmath}

\usepackage{ifthen} 

\newboolean{pdflatex}
\setboolean{pdflatex}{true} 
\newboolean{articletitles}
\setboolean{articletitles}{true} 
\newboolean{uprightparticles}
\setboolean{uprightparticles}{false} 
\usepackage{amssymb}
\usepackage{amsfonts}
\usepackage{upgreek}

\usepackage{hyperref}
\usepackage[all]{hypcap}

\def\lhcb {LHCb\xspace}
\def\ux85 {UX85\xspace}

\ifthenelse{\boolean{uprightparticles}}%
{

 \def\Ppsi        {\ensuremath{\uppsi}\xspace}

 \def\PDelta      {\ensuremath{\Delta}\xspace}                 
 \def\PXi      {\ensuremath{\Xi}\xspace}                 
 \def\PLambda      {\ensuremath{\Lambda}\xspace}                 
 \def\PSigma      {\ensuremath{\Sigma}\xspace}                 
 \def\POmega      {\ensuremath{\Omega}\xspace}                 
 \def\PUpsilon      {\ensuremath{\Upsilon}\xspace}                 
 
 \def\PB      {\ensuremath{\mathrm{B}}\xspace}                 
                  
 \def\PD      {\ensuremath{\mathrm{D}}\xspace}

 \def\PJ      {\ensuremath{\mathrm{J}}\xspace}                 
 \def\PK      {\ensuremath{\mathrm{K}}\xspace}

 \def\Pb      {\ensuremath{\mathrm{b}}\xspace}                 
 \def\Pc      {\ensuremath{\mathrm{c}}\xspace}

 \def\Pi      {\ensuremath{\mathrm{i}}\xspace}

}
{

 \def\Ppsi        {\ensuremath{\psi}\xspace}                 
                  
 \mathchardef\PDelta="7101
 \mathchardef\PXi="7104
 \mathchardef\PLambda="7103
 \mathchardef\PSigma="7106
 \mathchardef\POmega="710A
 \mathchardef\PUpsilon="7107
                  
 \def\PB      {\ensuremath{B}\xspace}                 
                  
 \def\PD      {\ensuremath{D}\xspace}

 \def\PJ      {\ensuremath{J}\xspace}                 
 \def\PK      {\ensuremath{K}\xspace}

 \def\Pb      {\ensuremath{b}\xspace}                 
 \def\Pc      {\ensuremath{c}\xspace}

 \def\Pi      {\ensuremath{i}\xspace}

}

\def\cquark    {\ensuremath{\Pc}\xspace}

\def\bquark    {\ensuremath{\Pb}\xspace}

\def\kaon  {\ensuremath{\PK}\xspace}
  \def\Kbar  {\kern 0.2em\overline{\kern -0.2em \PK}{}\xspace}

\def\Kz    {\ensuremath{\kaon^0}\xspace}
\def\Kzb   {\ensuremath{\Kbar^0}\xspace}
\def\KzKzb {\ensuremath{\Kz \kern -0.16em \Kzb}\xspace}
\def\Kp    {\ensuremath{\kaon^+}\xspace}
\def\Km    {\ensuremath{\kaon^-}\xspace}

\def\KpKm  {\ensuremath{\Kp \kern -0.16em \Km}\xspace}

  \def\Dbar    {\kern 0.2em\overline{\kern -0.2em \PD}{}\xspace}
\def\D       {\ensuremath{\PD}\xspace}

\def\Dz      {\ensuremath{\D^0}\xspace}
\def\Dzb     {\ensuremath{\Dbar^0}\xspace}
\def\DzDzb   {\ensuremath{\Dz {\kern -0.16em \Dzb}}\xspace}
\def\Dp      {\ensuremath{\D^+}\xspace}
\def\Dm      {\ensuremath{\D^-}\xspace}

\def\DpDm    {\ensuremath{\Dp {\kern -0.16em \Dm}}\xspace}

\def\B       {\ensuremath{\PB}\xspace}
  \def\Bbar    {\kern 0.18em\overline{\kern -0.18em \PB}{}\xspace}

\def\jpsi     {\ensuremath{{\PJ\mskip -3mu/\mskip -2mu\Ppsi\mskip 2mu}}\xspace}

  \def\Y#1S{\ensuremath{\PUpsilon{(#1S)}}\xspace}

\def\BR         {\BF}

\def\to                 {\ensuremath{\rightarrow}\xspace}

\def\AT#1     {\ensuremath{A_{\mathrm{T}}^{#1}}\xspace}           

\def\C#1      {\ensuremath{\mathcal{C}_{#1}}\xspace}                       
\def\Cp#1     {\ensuremath{\mathcal{C}_{#1}^{'}}\xspace}                    
\def\Ceff#1   {\ensuremath{\mathcal{C}_{#1}^{\mathrm{(eff)}}}\xspace}        
\def\Cpeff#1  {\ensuremath{\mathcal{C}_{#1}^{'\mathrm{(eff)}}}\xspace}       
\def\Ope#1    {\ensuremath{\mathcal{O}_{#1}}\xspace}                       
\def\Opep#1   {\ensuremath{\mathcal{O}_{#1}^{'}}\xspace}                    

\newcommand{\tev}{\ensuremath{\mathrm{\,Te\kern -0.1em V}}\xspace}
\newcommand{\gev}{\ensuremath{\mathrm{\,Ge\kern -0.1em V}}\xspace}
\newcommand{\mev}{\ensuremath{\mathrm{\,Me\kern -0.1em V}}\xspace}
\newcommand{\kev}{\ensuremath{\mathrm{\,ke\kern -0.1em V}}\xspace}
\newcommand{\ev}{\ensuremath{\mathrm{\,e\kern -0.1em V}}\xspace}
\newcommand{\gevc}{\ensuremath{{\mathrm{\,Ge\kern -0.1em V\!/}c}}\xspace}
\newcommand{\mevc}{\ensuremath{{\mathrm{\,Me\kern -0.1em V\!/}c}}\xspace}
\newcommand{\gevcc}{\ensuremath{{\mathrm{\,Ge\kern -0.1em V\!/}c^2}}\xspace}
\newcommand{\gevgevcccc}{\ensuremath{{\mathrm{\,Ge\kern -0.1em V^2\!/}c^4}}\xspace}
\newcommand{\mevcc}{\ensuremath{{\mathrm{\,Me\kern -0.1em V\!/}c^2}}\xspace}

\def\mum  {\ensuremath{\,\upmu\rm m}\xspace}

\def\gsim{{~\raise.15em\hbox{$>$}\kern-.85em
          \lower.35em\hbox{$\sim$}~}\xspace}
\def\lsim{{~\raise.15em\hbox{$<$}\kern-.85em
          \lower.35em\hbox{$\sim$}~}\xspace}

\def\PDF {PDF\xspace}

\def\tell1  {TELL1\xspace}
\def\ukl1   {UKL1\xspace}

\usepackage{cite}
\usepackage{mciteplus}
\begin{document}

\renewcommand{\thefootnote}{\fnsymbol{footnote}}
\setcounter{footnote}{1}


\begin{titlepage}
\pagenumbering{roman}

\vspace*{-1.5cm}
\centerline{\large EUROPEAN ORGANIZATION FOR NUCLEAR RESEARCH (CERN)}
\vspace*{1.5cm}
\hspace*{-0.5cm}
\begin{tabular*}{\linewidth}{lc@{\extracolsep{\fill}}r}
\ifthenelse{\boolean{pdflatex}}
{\vspace*{-2.7cm}\mbox{\!\!\!\includegraphics[width=.14\textwidth]{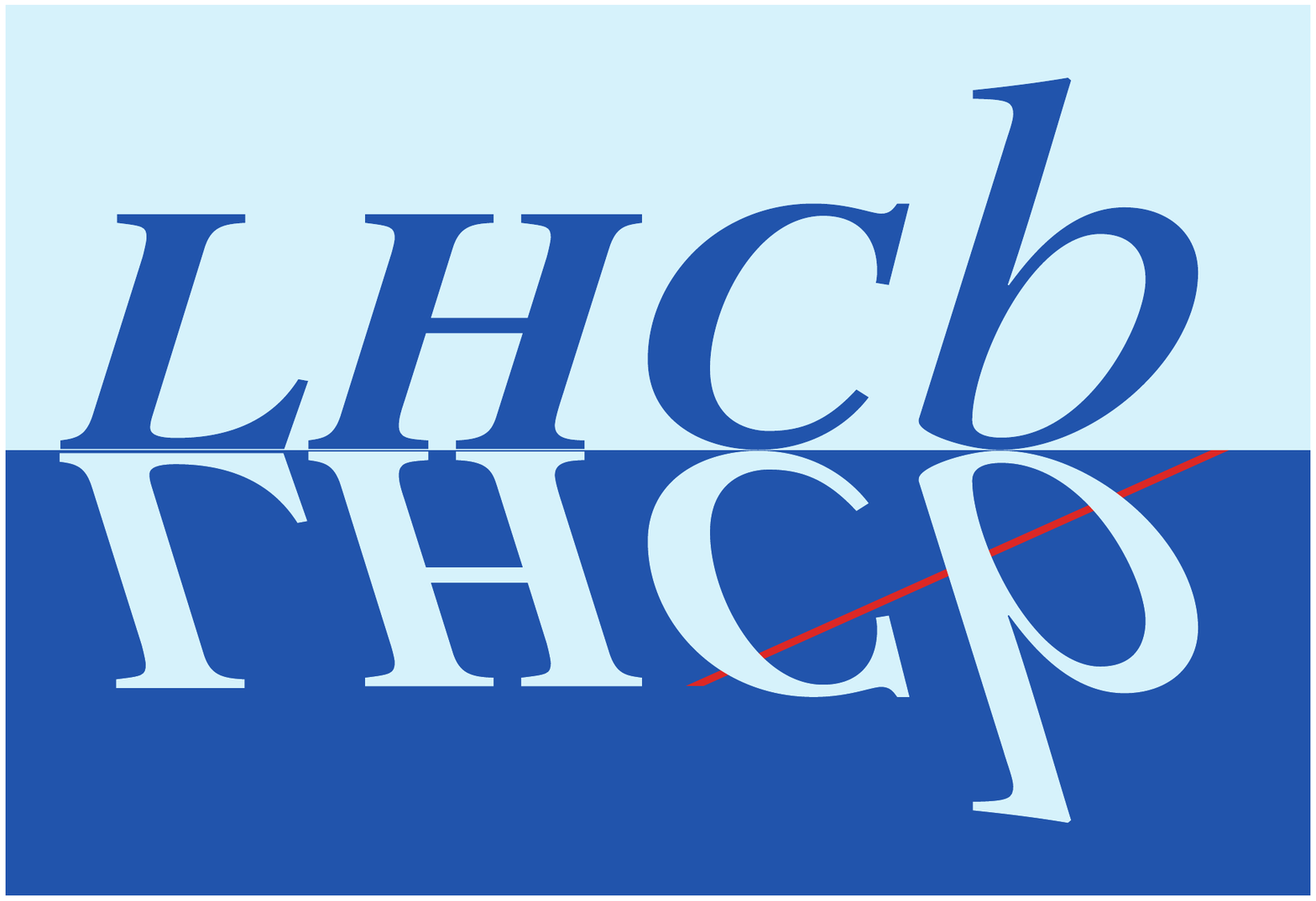}} & &}%
{\vspace*{-1.2cm}\mbox{\!\!\!\includegraphics[width=.12\textwidth]{lhcb-logo.eps}} & &}%
\\
 & & CERN-PH-EP-2012-004 \\  
 & & LHCb-PAPER-2011-033 \\  
 & & February 21, 2012 \\  
 & & \\
\end{tabular*}

\vspace*{4.0cm}

{\bf\boldmath\huge
\begin{center}
  Search for the $X(4140)$ state in $B^+\to J/\psi\phi K^+$ decays 
\end{center}
}

\vspace*{2.0cm}

\begin{center}
The LHCb collaboration
\footnote{Authors are listed on the following pages.}
\end{center}

\vspace{\fill}

\begin{abstract}
  \noindent
A search for the $X(4140)$ state in $B^+\to J/\psi\phi K^+$ decays is performed 
with 0.37 fb$^{-1}$ of $pp$ collisions at $\sqrt{s}=7$ TeV collected by the LHCb experiment. 
No evidence for this state is found,
in $2.4\,\sigma$ disagreement with a measurement by CDF.
An upper limit on its production rate is set, 
${{\cal B}(B^+\to X(4140) K^+)\times{\cal B}(X(4140)\to \jpsi \phi)}/{{\cal B}(B^+\to\jpsi \phi K^+)} <  0.07$ 
at 90\%\ confidence level.

\end{abstract}

\vspace*{2.0cm}

\begin{center}
  Submitted to Physical Review D Rapid Communications
\end{center}

\vspace{\fill}

\end{titlepage}


\newpage
\setcounter{page}{2}
\mbox{~}
\newpage

\begin{center}
{\large\bf The LHCb collaboration}
\end{center}
\begin{flushleft}
R.~Aaij$^{38}$, 
C.~Abellan~Beteta$^{33,n}$, 
B.~Adeva$^{34}$, 
M.~Adinolfi$^{43}$, 
C.~Adrover$^{6}$, 
A.~Affolder$^{49}$, 
Z.~Ajaltouni$^{5}$, 
J.~Albrecht$^{35}$, 
F.~Alessio$^{35}$, 
M.~Alexander$^{48}$
% ...

{\footnotesize \it
$ ^{1}$Centro Brasileiro de Pesquisas F\'{i}sicas (CBPF), Rio de Janeiro, Brazil\\
% ...
$ ^{5}$Clermont Universit\'{e}, Universit\'{e} Blaise Pascal, CNRS/IN2P3, LPC, Clermont-Ferrand, France\\
$ ^{6}$CPPM, Aix-Marseille Universit\'{e}, CNRS/IN2P3, Marseille, France\\
% ...
$ ^{33}$Universitat de Barcelona, Barcelona, Spain\\
$ ^{34}$Universidad de Santiago de Compostela, Santiago de Compostela, Spain\\
$ ^{35}$European Organization for Nuclear Research (CERN), Geneva, Switzerland\\
% ...
$ ^{38}$Nikhef National Institute for Subatomic Physics, Amsterdam, The Netherlands\\
% ...
$ ^{43}$H.H. Wills Physics Laboratory, University of Bristol, Bristol, United Kingdom\\
% ...
$ ^{48}$School of Physics and Astronomy, University of Glasgow, Glasgow, United Kingdom\\
$ ^{49}$Oliver Lodge Laboratory, University of Liverpool, Liverpool, United Kingdom\\
% ...
$ ^{n}$LIFAELS, La Salle, Universitat Ramon Llull, Barcelona, Spain\\
% ...
}
\bigskip
\end{flushleft}

\cleardoublepage

\renewcommand{\thefootnote}{\arabic{footnote}}
\setcounter{footnote}{0}



\pagestyle{plain} 
\setcounter{page}{1}
\pagenumbering{arabic}


%

\newlength{\figsize}
\setlength{\figsize}{0.7\hsize}
\def\bujkpp{B^+\to\jpsi K^+\pi^-\pi^+}
\def\bujkkk{B^+\to\jpsi K^+K^-K^+}
\def\bujphik{B^+\to\jpsi \phi K^+}
\def\buxk{B^+\to X(4140) K^+}
\def\BR{{\cal B}}
\def\DLL{{\rm DLL}}
\def\PDF{{\cal P}}
\def\NDOF{\hbox{\rm ndf}}
\def\coskj{\cos(K,\jpsi)}
\def\funone{{\cal F}^{\rm bkg}_1}
\def\funtwo{{\cal F}^{\rm bkg}_2}

In this article, results are presented from the search 
for the narrow $X(4140)$ resonance decaying to $\jpsi\phi$ 
using $\bujphik$ events\footnote{Charge-conjugate states are implied in this paper.} ($\jpsi\to\mu^+\mu^-$, $\phi\to K^+K^-$), 
in a data sample corresponding to an integrated luminosity of $0.37$~fb$^{-1}$  
collected in $pp$ collisions at the LHC at $\sqrt{s}=7$~TeV using the LHCb detector.
The CDF collaboration reported a 3.8$\,\sigma$ evidence for the $X(4140)$ state
(also referred to as $Y(4140)$ in the literature) in these decays
using $p\bar{p}$ data collected at the Tevatron ($\sqrt{s}=1.96$~TeV) \cite{Aaltonen:2009tz}. 
A preliminary update of the CDF analysis with 6.0~fb$^{-1}$ reported $115\pm12$ $\bujphik$ events and $19\pm6$ $X(4140)$ candidates 
leading to a statistical significance of more than 5$\,\sigma$ \cite{CDF}. 
The mass and width were determined to be $4143.4^{+2.9}_{-3.0}\pm0.6$~MeV and $15.3^{+10.4}_{-6.1}\pm2.5$~MeV, respectively\footnote{Units in which $c=1$ are used.}.  
The relative branching ratio was measured to be 
$\BR(\buxk)\times\BR(X(4140)\to\jpsi\phi)/\BR(\bujphik)=0.149\pm0.039\pm0.024$.

Charmonium states at this mass are expected to have much larger widths because of open flavour decay channels \cite{Brambilla:2010cs}.
Thus, their decay rate into the $\jpsi\phi$ mode, which is near the kinematic threshold, should be small and unobservable.
Therefore, the observation by CDF has triggered wide interest among model builders of exotic hadronic states.
It has been suggested that the $X(4140)$ resonance could be a molecular state 
\cite{Liu:2009ei,Branz:2009yt,Albuquerque:2009ak,Ding:2009vd,Zhang:2009st,Liu:2009pu,Wang:2009ry}, 
a tetraquark state \cite{Stancu:2009ka,Drenska:2009cd}, a hybrid state \cite{Mahajan:2009pj,Wang:2009ue} 
or a rescattering effect \cite{Liu:2009iw,Bugg:2011hf}. 
The Belle experiment found no evidence for the $X(4140)$ state in the $\gamma\gamma\to\jpsi\phi$ process,
which disfavoured the molecular interpretation \cite{Shen:2009vs}.  
The CDF data also suggested that there could be a second state at a mass of $4274.4^{+8.4}_{-6.4}\pm1.9$~MeV 
with a width of $32.3^{+21.9}_{-15.3}\pm7.6$~MeV \cite{CDF}.
In this case, the event yield was $22\pm8$ with $3.1\,\sigma$ significance.  
This observation has also received attention in the literature \cite{He:2011ed,Finazzo:2011he}.

The \lhcb detector~\cite{Alves:2008zz} is a single-arm forward
spectrometer covering the pseudo-rapidity range $2<\eta <5$, designed
for the study of particles containing \bquark or \cquark quarks. The
detector includes a high precision tracking system consisting of a
silicon-strip vertex detector surrounding the $pp$ interaction region,
a large-area silicon-strip detector located upstream of a dipole
magnet with a bending power of about $4{\rm\,Tm}$, and three stations
of silicon-strip detectors and straw drift-tubes placed
downstream. The combined tracking system has a momentum resolution
$\Delta p/p$ that varies from 0.4\% at 5\gevc to 0.6\% at 100\gevc,
and an impact parameter (IP) resolution of 20\mum for tracks with high
transverse momentum. Charged hadrons are identified using two
ring-imaging Cherenkov detectors. Photon, electron and hadron
candidates are identified by a calorimeter system consisting of
scintillating-pad and pre-shower detectors, an electromagnetic
calorimeter (ECAL) and a hadronic calorimeter (HCAL). Muons are identified by a muon
system (MUON) composed of alternating layers of iron and multiwire
proportional chambers. 
The MUON, ECAL and HCAL
provide the capability of first-level hardware triggering.
The single and dimuon hardware triggers provide good efficiency for $\bujphik$, $\jpsi\to\mu^+\mu^-$ events. 
Events passing the hardware trigger are read out and sent to an event filter farm for further processing. 
Here, a software based two-stage trigger reduces the rate from 1~MHz to about 3~kHz.
The most efficient software triggers \cite{LHCb-PUB-2011-016} 
for this analysis require a charged track with transverse momentum ($p_{\rm T}$) of more than $1.7$~GeV 
($p_{\rm T}>1.0$~GeV if identified as muon) 
and with an IP to any primary $pp$-interaction vertex (PV) larger than $100$~$\mu$m.
A dimuon trigger requiring $p_{\rm T}(\mu)>0.5$~GeV, large dimuon mass, $M(\mu^+\mu^-)>2.7$~GeV,
and with no IP requirement complements the single track triggers.
At final stage, we either 
require a $\jpsi\to\mu^+\mu^-$ candidate with $p_{\rm T}>1.5$~GeV
or a muon-track pair with significant IP.
  
In the subsequent offline analysis, 
$\jpsi\to\mu^+\mu^-$ candidates are selected with the following criteria: $p_{\rm T}(\mu)>0.9$~GeV, 
$\chi^2$ per degree of freedom of the two muons forming a common vertex, $\chi^2_{\rm vtx}(\mu^+\mu^-)/\NDOF<9$,
and a mass window $3.04< M(\mu^+\mu^-)<3.14$~GeV.   
We then find 
$K^+K^-K^+$ combinations consistent with originating from a common vertex with 
$\chi^2_{\rm vtx}(K^+K^-K^+)/\NDOF<9$.
Every charged track with $p_{\rm T}>0.25$~GeV,
missing all PVs by at least 3 standard deviations ($\chi^2_{\rm IP}(K)>9$)
and classified more likely to be a kaon than a pion according to the particle identification system,
is considered a kaon candidate. 
A five-track $\jpsi K^+K^-K^+$ vertex is formed ($\chi^2_{\rm vtx}(\jpsi K^+K^-K^+)/\NDOF<9$).
This $B^+$ candidate is required to have $p_{\rm T}>4.0$~GeV and a decay time 
as measured with respect to the PV of at least $0.25$ ps.
When more than one PV is reconstructed, the one
that gives the smallest IP significance for the $B^+$ candidate is chosen.
The invariant mass of a $\mu^+\mu^- K^+K^-K^+$ combination is evaluated after 
the muon pair is constrained to the $\jpsi$ mass, and all final state particles are constrained 
to a common vertex. 

Further background suppression 
is provided by a likelihood ratio. 
In the case of uncorrelated input variables this provides 
the most efficient discrimination between signal and background. 
The overall likelihood is a product of 
probability density functions, $\PDF(x_i)$ (PDFs),  
for the four sensitive variables ($x_i$): 
smallest $\chi^2_{\rm IP}(K)$ among the kaon candidates,
$\chi^2_{\rm vtx}(\jpsi K^+K^-K^+)/\NDOF$,
the pointing of the $B^+$ candidate to the closest primary vertex, $\chi^2_{\rm IP}(B)$,
and the cosine of the largest opening angle between the $\jpsi$ and kaon candidates 
in the plane transverse to the beam.
The latter peaks towards $+1$ for the signal as the $B^+$ meson has a high transverse momentum.
Backgrounds combining particles from two different $B$ mesons peak at $-1$.
Backgrounds including other random combinations are uniformly distributed.  
The signal PDFs, $\PDF_{\rm sig}(x_i)$, 
are obtained from the Monte Carlo simulation (MC) of 
$\bujkkk$ decays.
The background PDFs, $\PDF_{\rm bkg}(x_i)$, are obtained from the data 
candidates with $\jpsi K^+K^-K^+$ invariant mass between 5.6 and 6.4~GeV (far-sideband).
A logarithm of the ratio of the signal and background PDFs is formed:
$\DLL_{\rm sig/bkg} = -2 \sum_i^4 \ln(\PDF_{\rm sig}(x_i)/\PDF_{\rm bkg}(x_i))$.
A requirement on the log-likelihood ratio, $\DLL_{\rm sig/bkg}<-1$, has been chosen by 
maximizing $N_{\rm sig}/\sqrt{N_{\rm sig}+N_{\rm bkg}}$, 
where $N_{\rm sig}$ is the expected $\bujkkk$ signal yield and
the $N_{\rm bkg}$ is the background yield in the $B^+$ peak region ($\pm2.5\,\sigma$).
The absolute normalization of $N_{\rm sig}$ and $N_{\rm bkg}$ comes from a fit 
to the $\jpsi\phi K$ invariant mass distribution
with $\DLL_{\rm sig/bkg}<0$, while their dependence on 
the $\DLL_{\rm sig/bkg}$ requirement comes from the signal simulation 
and the far-sideband, respectively. 

\begin{figure}[htbp]
  \begin{center}
  \ifthenelse{\boolean{pdflatex}}{
    \includegraphics*[width=\figsize]{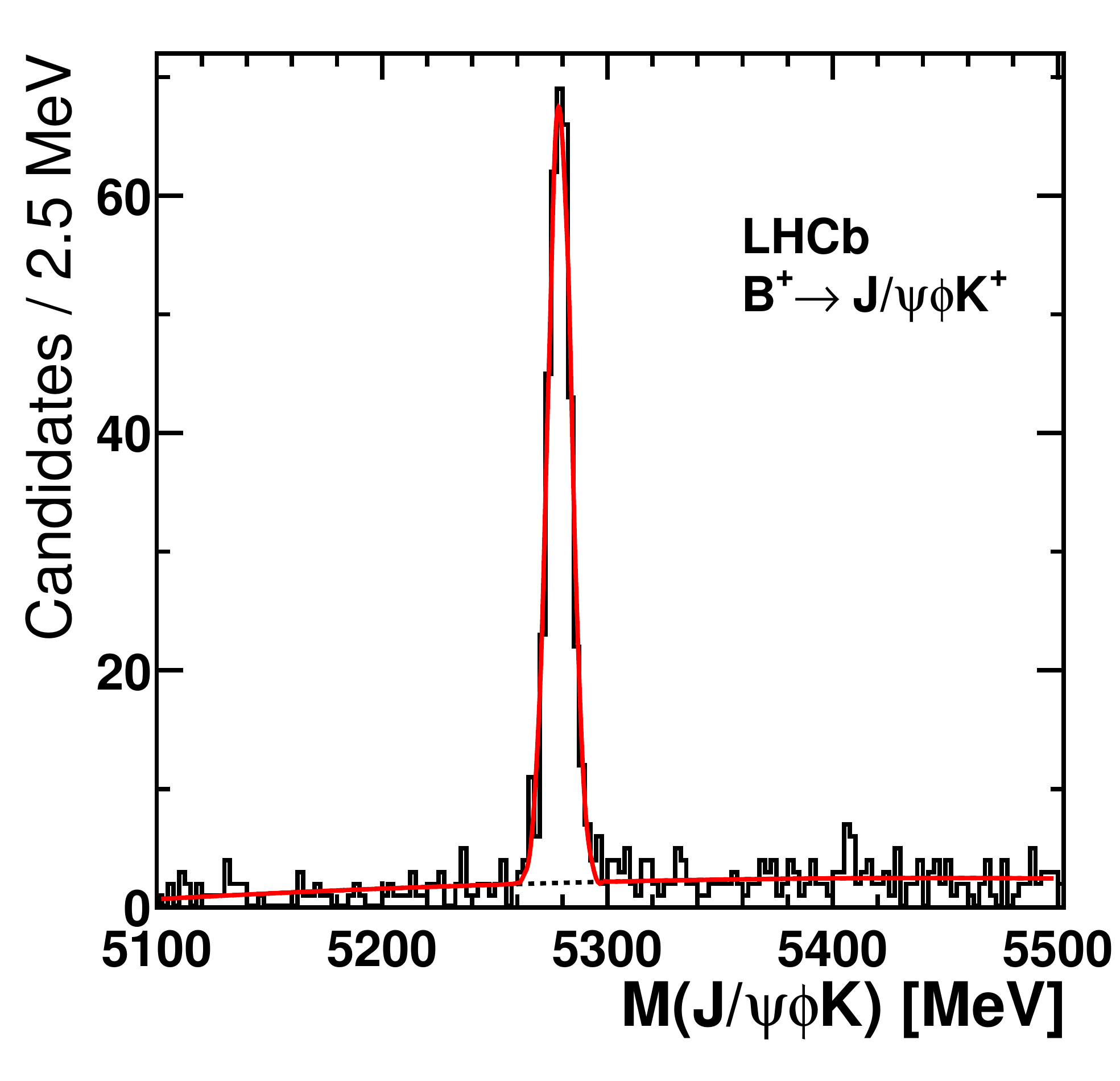}
   }{
    \includegraphics*[width=\figsize]{figx_jkkkmphi.eps}
   } 
  \end{center}
   \vskip-0.3cm
   \caption{\small
    Mass distribution for $\bujphik$ candidates in the data 
    after the $\pm15$~MeV $\phi$ mass requirement.
    The fit of a Gaussian signal with a quadratic background (dashed line) is
    superimposed (solid red line).}
  \label{fig:jkkkmphi}
\end{figure}

\begin{figure}[htbp]
  \begin{center}
  \ifthenelse{\boolean{pdflatex}}{
    \includegraphics*[width=\figsize]{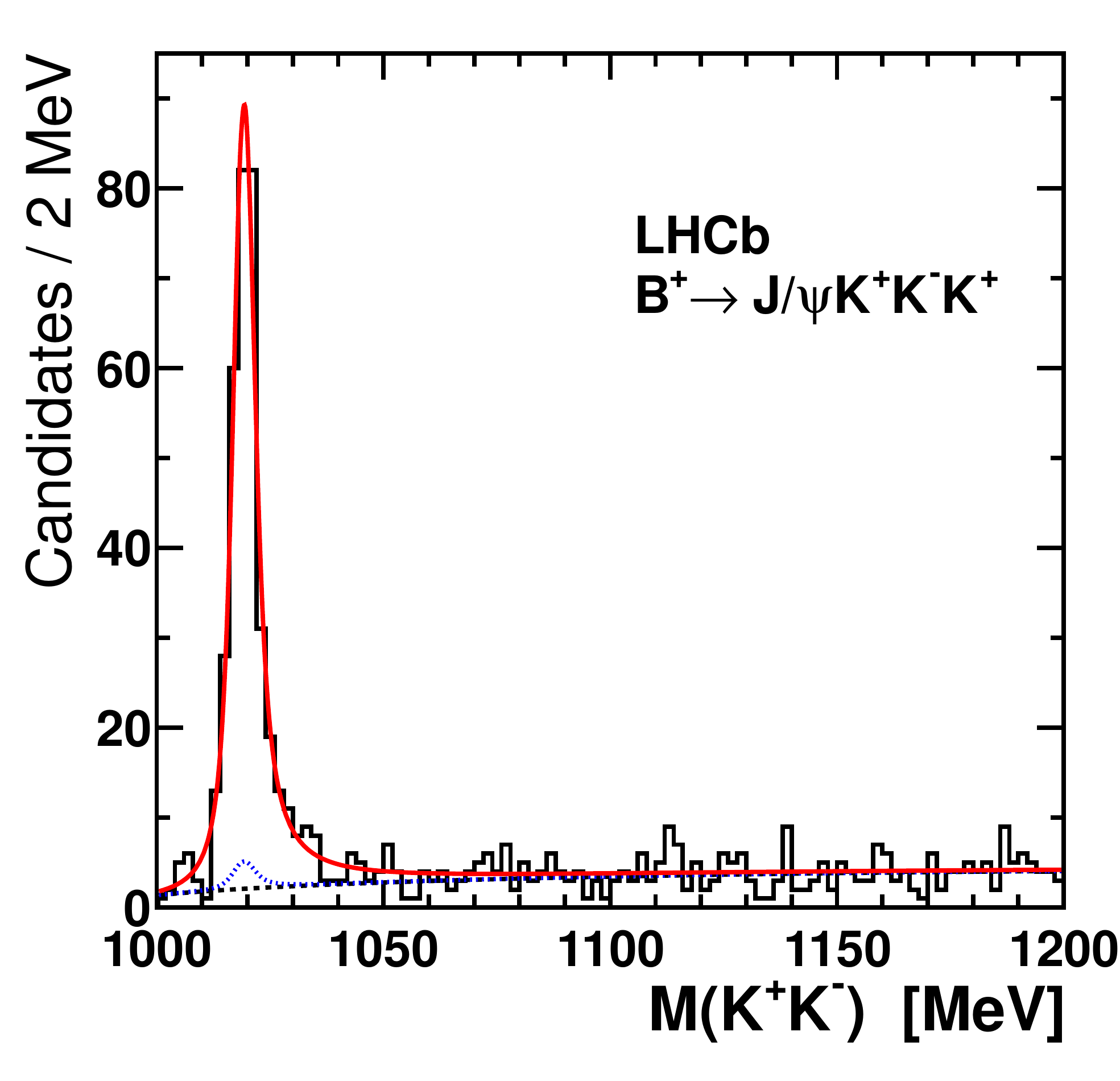}
   }{
    \includegraphics*[width=\figsize]{figx_kkmfit.eps}
   } 
  \end{center}
  \vskip-0.3cm
 \caption{\small 
    Invariant $M(K^+K^-)$ mass distribution selecting $\bujkkk$ events in the $\pm2.5\,\sigma$ region around the $B^+$ mass peak.  
    The dashed line shows the two-body phase-space contribution. 
    The small blue 
    dotted $\phi$ peak on top of it illustrates the amount of the background $\phi$ mesons estimated from the fit to the $B^+$ mass 
    near-sidebands.} 
  \label{fig:mkkfit}
\end{figure}

The $\jpsi\phi K$ invariant mass distribution, with a requirement that at least one $K^+K^-$ combination 
has an invariant mass within $\pm15$~MeV of the $\phi$ mass, is shown in Fig.~\ref{fig:jkkkmphi}. 
A fit to a Gaussian and a quadratic function in the range $5.1-5.5$~GeV results in $346\pm20$ $B^+$ events 
with a mass resolution of $5.2\pm0.3$~MeV. 
Alternatively 
requiring the invariant mass $M(\jpsi K^+K^-K^+)$ to be within $\pm2.5$ standard deviations 
of the observed $B^+$ peak position,
we fit the $M(K^+K^-)$ mass distribution (two combinations per event) 
using a binned maximum likelihood fit 
with a P-wave relativistic Breit-Wigner representing the $\phi(1020)$ and a
two-body phase-space distribution to represent combinatorial background,
both convolved with a Gaussian mass resolution.
The $\phi$ resonance width is fixed to the PDG value ($4.26$~MeV) \cite{PDG}. 
The $M(K^+K^-)$ mass distribution is displayed in Fig.~\ref{fig:mkkfit}
with the fit results overlaid. 
The fitted parameters are the $\phi$ yield, 
the $\phi$ mass ($1019.3\pm0.2$~MeV),
the background yield and the mass resolution 
($1.4\pm0.3$~MeV).
Replacing the two-body phase-space function by a third-order polynomial does not change the results.
In order to subtract a non-$B$ contribution, we fit the $M(K^+K^-)$ distribution from the $B$ mass near-sidebands 
(from $4$ to $14$ standard deviations on either side) leaving only the $\phi$ yield and the two-body phase-space background yield as
free parameters. After scaling to the signal region, this leads to $14\pm3$ background events.
The background subtracted $\bujphik$ yield ($N_{\bujphik}$) is $382\pm22$ events.

\begin{figure}[htbp]
  \begin{center}
  \ifthenelse{\boolean{pdflatex}}{
    \includegraphics*[width=\figsize]{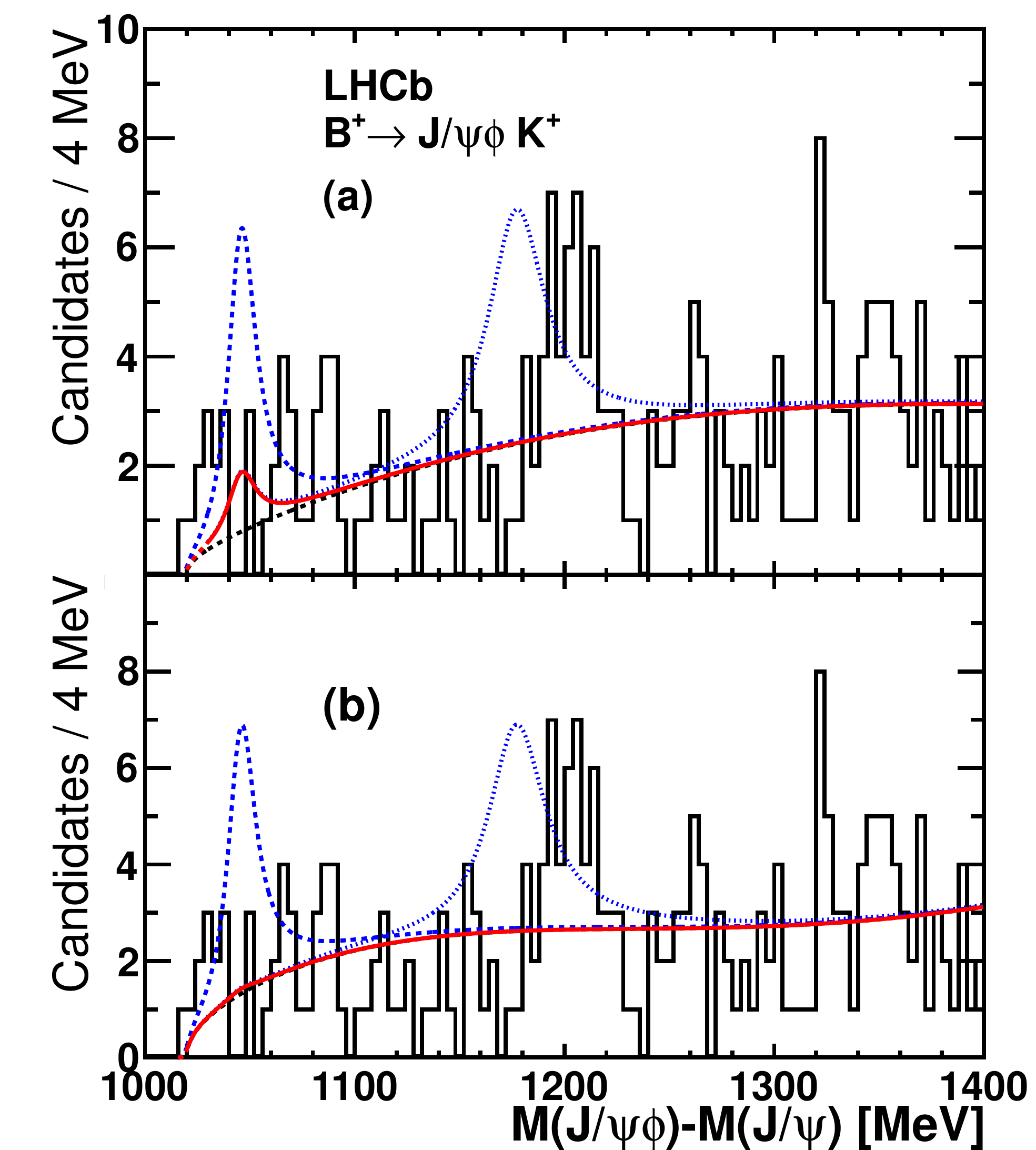}
   }{
    \includegraphics*[width=\figsize]{figx_jkkjfit.eps}
   } 
  \end{center}
  \vskip-0.3cm
  \caption{\small  
          Distribution of the mass difference $M(\jpsi\phi)-M(\jpsi)$ for the $\bujphik$ 
          in the $B^+$ ($\pm2.5\,\sigma$) and $\phi$ ($\pm15$~MeV) mass windows.           
                  Fit of $X(4140)$ signal on top of a smooth background is superimposed (solid red line). 
                  The dashed blue (dotted blue) line on top illustrates the expected $X(4140)$ ($X(4274)$) signal yield from the 
		  CDF measurement \cite{CDF}.                  
                  The top and bottom plots differ by the background function (dashed black line)
		  used in the fit:
                  (a) an efficiency-corrected three-body phase-space ($\funone$);                      
                  (b) a quadratic function multiplied by the efficiency-corrected three-body phase-space factor ($\funtwo$).  		  
		  The fit ranges are 1030--1400 and 1020--1400~MeV, respectively.
                  } 
  \label{fig:jkkjfit}
\end{figure}

\begin{figure}[htbp]
  \begin{center}
  \ifthenelse{\boolean{pdflatex}}{
    \includegraphics*[width=\figsize]{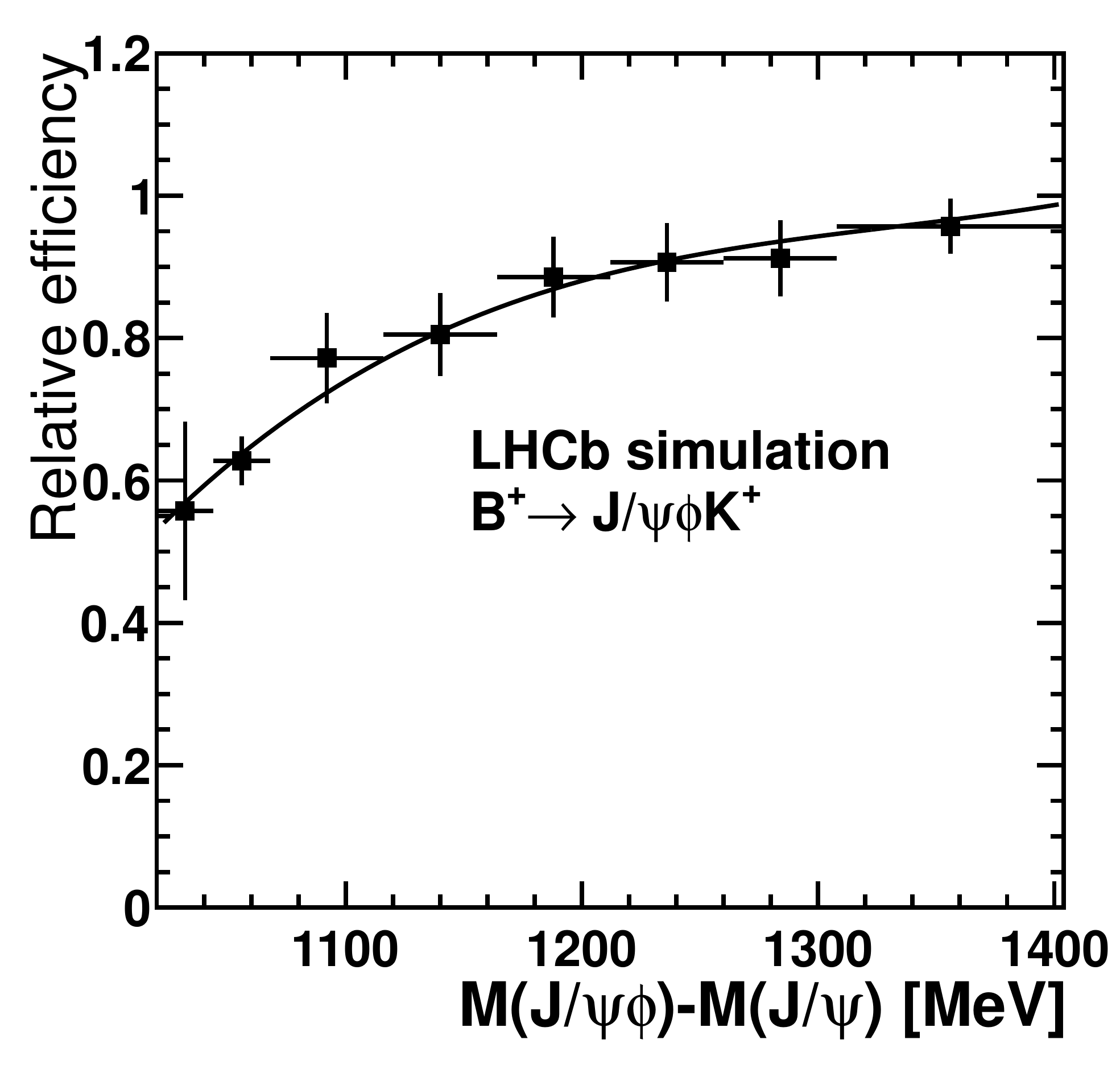}
   }{
    \includegraphics*[width=\figsize]{figx_effjkkj.eps}
   } 
  \end{center}
  \vskip-0.3cm
   \caption{\small 
    Efficiency dependence on $M(\jpsi\phi)-M(\jpsi)$ as determined from the simulation (points with error bars).
    The efficiency is normalized with respect to the efficiency of the $\phi$ signal fit to the $\bujphik$ events 
    distributed according to the phase-space model. 
    A cubic polynomial was fitted to the simulated data (superimposed).} 
  \label{fig:eff}
\end{figure}

To search for the $X(4140)$  state, we select events within $\pm15$~MeV of the $\phi$ mass.
According to the fit to the $M(K^+K^-)$ distribution this requirement is 85\%\ efficient. 
Figure~\ref{fig:jkkjfit} shows the mass difference $M(\jpsi\phi)-M(\jpsi)$ 
distribution (no $\jpsi$ or $\phi$ mass constraints have been used).
No narrow structure is observed near the threshold. 
We employ the fit model used by CDF \cite{CDF} to quantify the compatibility of the two measurements. 
The data are fitted with a spin-zero relativistic Breit-Wigner shape 
together with a three-body phase-space function ($\funone$), both convolved with the detector resolution.
The efficiency dependence is extracted from simulation (Fig.~\ref{fig:eff}) and applied 
as a shape correction to the three-body phase-space and the Breit-Wigner function.
The mass and width of the $X(4140)$ peak are fixed to the central values obtained by the CDF collaboration.
The mass-difference resolution was determined from the $\buxk$ simulation to be $1.5\pm0.1$~MeV. 
A binned maximum likelihood fit of the signal and background yields 
is shown in Fig.~\ref{fig:jkkjfit}(a).
The region above 1400~MeV is excluded since it is more likely to contain 
non $\bujphik$ backgrounds.
By excluding also the region below 1030~MeV, where  
the three-body phase-space and signal yields are very small
($0.5\%$ and $3.5\%$ of the yields included in the fit, respectively), we make
our results less vulnerable to possible small contributions from the other sources.   
The fit shown in Fig.~\ref{fig:jkkjfit}(a) 
gives a $X(4140)$ yield of $6.9\pm4.9$ events. 
Fitting the second state at a mass of $4274.44$~MeV and with a width of $32.3$~MeV \cite{CDF}
does not affect the $X(4140)$ yield.
Reflections of $K\phi$ resonances \cite{Frame:1985ka,Armstrong:1982tw} 
and possible broad $J/\psi\phi$ resonances can also contribute near and under 
the narrow $X(4140)$ resonance.   
To explore the sensitivity of our results to the assumed background shape, we also fit the data
in the $1020-1400$~MeV range with a quadratic function multiplied by 
the efficiency-corrected three-body phase-space factor ($\funtwo$) 
to impose the kinematic threshold.  
The preferred  value of the $X(4140)$ yield is $0.6$ events with a positive error of $7.1$ events.  
This fit is shown in Fig.~\ref{fig:jkkjfit}(b).

A similar fit was performed to simulated $\buxk$ data to estimate the efficiency for this channel.
The efficiency ratio between this fit and the $\phi$ signal fit to the $\bujphik$ events 
distributed according to the phase-space model, 
$\epsilon(\buxk, X(4140)\to \jpsi \phi)/\epsilon(\bujphik)$,
was determined to be $0.62\pm0.04$ and includes the efficiency of 
the $\phi$ mass window requirement.
Using our $\bujphik$ yield multiplied by this efficiency ratio and 
by the CDF value for $\BR(\buxk)/\BR(\bujphik)$ \cite{CDF}, leads to a prediction
that we should have observed $35\pm9\pm6$ events, where the first uncertainty is statistical from the CDF data 
and the second includes both the CDF and LHCb systematic uncertainties. 
Given the $B^+$ yield and relative efficiency,
our sensitivity to the $X(4140)$ signal is a factor of two better than that of the CDF.
The central value of this estimate is shown as a dashed line in Fig.~\ref{fig:jkkjfit}.
Taking the statistical and systematic errors from both experiments into account,
our results disagree with the CDF observation 
by 2.4$\,\sigma$ (2.7$\,\sigma$) when using $\funone$ ($\funtwo$) background shapes. 

Since no evidence for the $X(4140)$ state is found, we set an upper limit on its production.
Using a Bayesian approach, we
integrate the fit likelihood determined as a function of the $X(4140)$ yield
and find an upper limit on the number of signal events of $16$ ($13$) 
at 90\% 
confidence level (CL)
for the two background shapes. 
Dividing the least stringent limit on the signal yield 
by the $\bujphik$ yield and $\epsilon(\buxk)/\epsilon(\bujphik)$ gives
a limit on $\BR(\buxk)\times\BR(X(4140)\to\jpsi\phi)/\BR(\bujphik)$. 
The systematic uncertainty on $\epsilon(\buxk)/\epsilon(\bujphik)$ is 6\%.
This uncertainty includes the statistical error from the simulation as well as 
the observed differences in track reconstruction efficiency 
between the simulation and data measured with
the inclusive $\jpsi\to\mu^+\mu^-$ signal.  
Fit systematics related to the detector resolution and the uncertainty in the shape of the 
efficiency dependence on the $\jpsi\phi$ mass were also studied and found to be small.
We multiply our limit by 1.06 and obtain at 90\%~CL
\begin{equation*}  
\frac{\BR(\buxk)\times\BR(X(4140)\to \jpsi \phi)}{\BR(\bujphik)} <  0.07.
\end{equation*}

We also set an upper limit on the $X(4274)$ state suggested by the CDF collaboration \cite{CDF}.
The fit with $\funone$ background shape gives $3.4^{+6.5}_{-3.4}$ events at this mass.  
The fit with the $\funtwo$ background shape gives zero signal events with a positive error of $10$. 
Integration of the fit likelihoods gives $<24$  and $<20$ events at 90\% CL, respectively. 
The relative efficiency at this mass 
is $\epsilon(B^+\to X(4274) K^+, X(4274)\to \jpsi \phi)/\epsilon(\bujphik)=0.86\pm0.10$. 
The least stringent limit on the signal events yields an upper limit of
\begin{equation*}  
\frac{\BR(\B^+\to X(4274) K^+)\times\BR(X(4274)\to \jpsi \phi)}{\BR(\bujphik)} <  0.08 
\end{equation*}
at~90\%~CL, which includes the systematic uncertainty.
CDF did not provide a measurement of this ratio of branching fractions. 
Assuming the efficiency is similar for the $X(4274)$ and
$X(4140)$ resonances, their $X(4274)$ event yield corresponds to
$\BR(\B^+\to X(4274) K^+)\times\BR(X(4274)\to\jpsi\phi)/\BR(\bujphik)=0.17\pm0.06$ (statistical uncertainty only).
Scaling to our data, we should have observed $53\pm19$ $X(4274)$ events, which is illustrated 
in Fig.~\ref{fig:jkkjfit}.    


In summary, 
the most sensitive search for the narrow $X(4140)\to\jpsi\phi$ state just above the kinematic threshold
in $\bujphik$ decays has been performed using 0.37~fb$^{-1}$ of data collected with the LHCb detector.
We do not confirm the existence of such a state. 
Our results disagree at the $2.4\,\sigma$ level with the CDF measurement.
An upper limit
on ${\BR(\buxk)\times\BR(X(4140)\to \jpsi \phi))/}$ ${\BR(\bujphik)}$ 
of $<0.07$ at 90\%~CL is set.

\section*{Acknowledgments}

We express our gratitude to our colleagues in the CERN accelerator
departments for the excellent performance of the LHC. We thank the
technical and administrative staff at CERN and at the LHCb institutes,
and acknowledge support from the National Agencies: CAPES, CNPq,
FAPERJ and FINEP (Brazil); CERN; NSFC (China); CNRS/IN2P3 (France);
BMBF, DFG, HGF and MPG (Germany); SFI (Ireland); INFN (Italy); FOM and
NWO (The Netherlands); SCSR (Poland); ANCS (Romania); MinES of Russia and
Rosatom (Russia); MICINN, XuntaGal and GENCAT (Spain); SNSF and SER
(Switzerland); NAS Ukraine (Ukraine); STFC (United Kingdom); NSF
(USA). We also acknowledge the support received from the ERC under FP7
and the Region Auvergne.

\bibliographystyle{LHCb}
\bibliography{main}
\end{document}